\documentclass[10pt, letterpaper]{IEEEtran}

\usepackage{cite}
\usepackage[cmex10]{amsmath}
\usepackage{algorithmic}
\usepackage{array}
\usepackage{url}
\usepackage{footnote}
\usepackage{float}
\usepackage[utf8]{inputenc}

\usepackage[T1]{fontenc}
\usepackage[ruled,vlined]{algorithm2e}
\usepackage{ifthen}
\usepackage{graphicx}
\usepackage{comment}
\usepackage{amsthm}
\usepackage{amsmath}
\usepackage{amsmath,amssymb,amsfonts}
\usepackage{nicefrac}
\usepackage{algorithmic}
\usepackage{bm}
\usepackage{multirow}
\usepackage{mathtools}
\usepackage{tikz}
\usetikzlibrary{calc,math}
\usepackage[framemethod=TikZ]{mdframed}
\usepackage{pgfplots}
\pgfplotsset{compat=1.16}
\usepackage{hyperref}
\hypersetup{colorlinks=true, unicode=true, linkcolor=[rgb]{0.10,0.05,0.67}, citecolor=[rgb]{0.10,0.05,0.67}, filecolor=[rgb]{0.10,0.05,0.67}, urlcolor=[rgb]{0.10,0.05,0.67}}

\newtheorem{lemma}{Lemma}

\newtheorem{corollary}{Corollary}

\DeclarePairedDelimiter{\floor}{\lfloor}{\rfloor}
\definecolor{DarkGreen}{rgb}{0.1,0.5,0.1}
\definecolor{DarkRed}{rgb}{0.5,0.1,0.1}
\definecolor{DarkBlue}{rgb}{0.1,0.1,0.5}
\definecolor{DarkPurple}{rgb}{0.5,0.2,0.5}
\definecolor{DarkTurquoise}{rgb}{0.1,0.5,0.5}

\makeatletter
\renewcommand*{\eqref}[1]{%
  \hyperref[{#1}]{\textup{\tagform@{\ref*{#1}}}}%
}
\makeatother

\begin{document}

% \title{On the Benefits of Coding in 6G Networks}
\title{Revisiting the Interface between Error and Erasure Correction in Wireless Standards\thanks{This manuscript builds on and extends preliminary concepts introduced in our prior works \cite{esfahanizadeh2024benefits,landon2024enhancing}. This work has been accepted for publication in \emph{IEEE Journal on Selected Areas in Communications (JSAC)}. © IEEE. Personal use of this material is permitted. Permission from IEEE must be obtained for all other uses.}}%Opportunities for network coding in standardization}

% Error correcting coding in 5G and before, why not erasure correction?
% Get figure from Muriel
% 
\author{
\IEEEauthorblockN{Vipindev Adat Vasudevan$^{\dagger}$, Homa Esfahanizadeh$^{\ddagger}$, Benjamin D. Kim$^{+}$, Laura Landon$^{\dagger}$, \\Alejandro Cohen$^{*}$, and Muriel M\'edard$^\dagger$}\\
\IEEEauthorblockA{
$^\dagger$Massachusetts Institute of Technology (MIT), Cambridge, USA, Emails: \{vipindev, llandon9, medard\}@mit.edu
}\\
\IEEEauthorblockA{
$^\ddagger$Nokia Bell Labs, Murray Hill, NJ, USA, Email: homa.esfahanizadeh@nokia-bell-labs.com
}\\
\IEEEauthorblockA{
$^+$University of Illinois Urbana-Champaign (UIUC), Champaign, Illinois, USA, Email: bdkim4@illinois.edu
}\\
\IEEEauthorblockA{
$^*$Technion-Israel Institute of Technology, Haifa, Israel, Email: alecohen@technion.ac.il
}
%\vspace{-0.3cm}

}

\maketitle

\begin{abstract}
Modern 5G communication systems implement a combination of error correction and feedback-based erasure correction (HARQ/ARQ) as reliability mechanisms, which can introduce substantial delay and resource inefficiency. We propose forward erasure correction using network coding as a more delay-efficient alternative. We present a mathematical characterization of network delay for existing reliability mechanisms and network coding. Through simulations in a network slicing environment, we demonstrate that network coding not only improves the in-order delivery delay and goodput for the applications utilizing the slice, but also benefits other applications sharing the network by reducing resource utilization for the coded slice. Our analysis and characterization point towards ideas that require attention in the 6G standardization process. These findings highlight the need for greater modularity in protocol stack design that enables the integration of novel technologies in future wireless networks.
\end{abstract}

\section{Introduction}

% Solving low delay requirements, issues with current trends
The fifth generation (5G) of wireless networks marked a fundamental shift in the design of communication systems. While previous generations primarily focused on incremental improvements in data rates with minimal changes to the underlying infrastructure, 5G introduced a transformative architectural overhaul. It enabled a diverse set of use cases, including massive Machine-Type Communications (mMTC) \cite{3gpp.22.861}, Ultra-Reliable Low-Latency Communications (URLLC) \cite{3gpp.22.862}, and enhanced Mobile Broadband (eMBB) \cite{3gpp.22.863}. However, as demands continue to grow and new technologies emerge, the next-generation of wireless networks is expected to build upon the foundation laid by 5G while addressing its limitations and new challenges. Emerging applications in the sixth-generation network (6G) such as extended reality (XR) and virtual reality (VR) will require ultra-low latency and extreme reliability \cite{pourkabirian2024vision,10879074,10859271}. Developments in the field of artificial intelligence (AI) are set to play a central role in optimizing network operations, enhancing efficiency, and enabling intelligent resource management. Furthermore, 6G envisions the inclusion of non-terrestrial networks, integrated sensing and communication, and an open, cooperative network environment among service providers. The evolution toward 6G reflects a broader shift toward intelligent, adaptive networks capable of meeting increasingly diverse and dynamic communication requirements, driving innovation across multiple domains and ensuring that future wireless networks can keep pace with the ever-growing demands of users and industries.

The IMT 2030 future trends and framework documents \cite{wp5d2022future,wp5d2023m,singh2025towards} outline the vision and objectives for 6G, emphasizing its role in meeting the ever-evolving demands of wireless communication. A key target of 6G is to enable Hyper Reliable Low Latency Communication (HRLLC), ensuring seamless connectivity for critical applications with stringent performance requirements. Achieving this goal will necessitate more efficient utilization of shared resources and the widespread adoption of virtual network architectures capable of supporting heterogeneous services. To achieve the expected quality of service for different use cases using the same network infrastructure, dedicated virtual networks will be assigned. This idea of \textit{network slicing} will be even more crucial in the 6G era. However, just providing dedicated virtual paths for communication will not suffice to meet the requirements of reliable communication with an extremely low latency budget. 

\begin{figure*}[!t]
  \centering
  \includegraphics[width=0.95\linewidth]{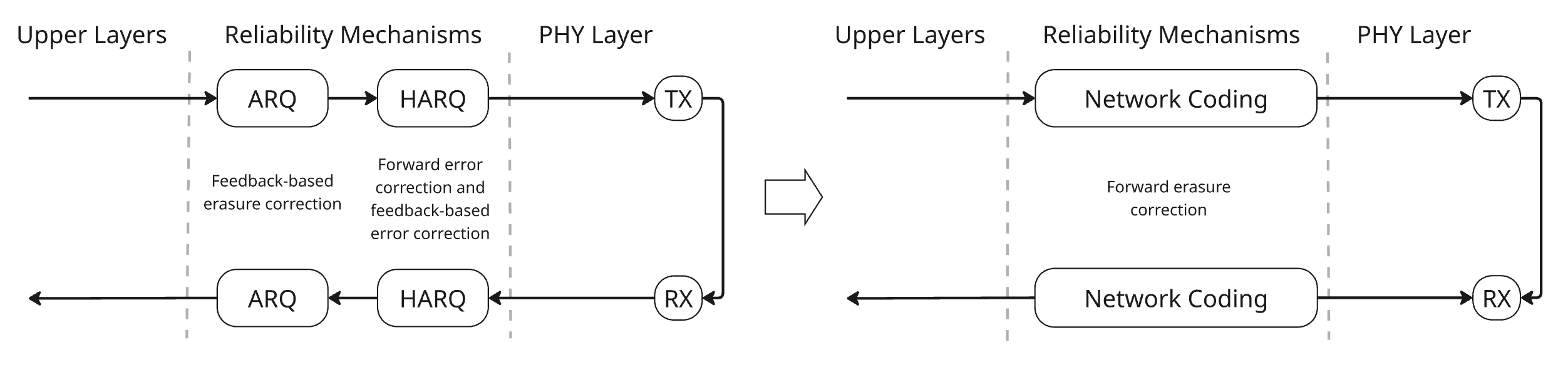}
  \caption{HARQ in the MAC layer and ARQ in the RLC layer of the current 5G protocol stack perform error correction based on feedback. Using a forward erasure correction (FEC) scheme, such as network coding, can reduce the delay. Replacing multi-layer, feedback-based error correction with a FEC scheme can significantly reduce delay in future wireless systems.} 
  \label{fig_intro}
\end{figure*}

Current wireless networks primarily rely on feedback-based reliability mechanisms, such as Automatic Repeat reQuest (ARQ) and Hybrid ARQ (HARQ), to correct packet losses. ARQ performs erasure correction by retransmitting entire lost frames upon receiving negative acknowledgments, whereas HARQ combines ARQ with error correction by retransmitting additional redundancy bits to recover from errors.  However, these mechanisms often struggle to balance reliability with service time, particularly under stringent latency constraints. In 5G systems, HARQ operates at the Medium Access Control (MAC) layer using an incremental redundancy (IR) scheme with Low-Density Parity-Check (LDPC) codes, combined with adaptive modulation and coding scheme (MCS) selection. Typically, the MCS is chosen to yield a 10\% target Block Error Rate (BLER) on the initial transmission \cite{kim2012network, 3gpp.38.214}. When decoding fails, additional parity bits are retransmitted. If HARQ retransmissions are insufficient, the upper-layer ARQ at the Radio Link Control (RLC) layer handles recovery by retransmitting the entire packet. While HARQ introduces limited forward error correction, approximately 10\% of packets still require retransmission, introducing round-trip time (RTT) delays that degrade latency performance. Furthermore, when the defined number of HARQ transmissions fails, upper-layer ARQ transmissions will kick in, discarding the transport block altogether. Thus, the current multi-layer reliability mechanisms based on feedback and error correction introduce redundancy as well as feedback-induced delay. Except for the packets that are received correctly at the first HARQ transmission that provides a forward error correction, each transport block incurs an RTT delay, undermining the latency and reliability targets of URLLC use case.  Figure \ref{fig_intro} highlights this inefficiency, motivating our proposed solution: a forward erasure correction (FEC) approach that eliminates the need for feedback-driven retransmissions, reducing delay and simplifying the reliability mechanism. Specifically, we employ random linear network coding (RLNC), a method within the broader class of erasure correction techniques, that transmits additional coded packets in advance to compensate for the losses. This changes the protocol to be proactive against losses than the current reactive mechanism. Furthermore, the RLNC approach eliminates the need for per packet feedback and allows block acknowledgments. This new approach also allows additional coded packets to be transmitted based on the block acknowledgment in the extreme cases where the apriori FEC packets were not enough to compensate for losses.

FEC coding schemes, such as network coding-based approaches \cite{ho2006random,cloud2015coded,cohen2020adaptive,cohen2021adaptive,cohen2021bringing}, are being explored as promising solutions to address the trade-offs between reliability, latency, and throughput in 6G and beyond. As standardization efforts for 6G technologies progress, the adoption of advanced FEC techniques that were not included in previous-generation standards is also considered \cite{adjakple2025user,geiselhart20236g}. In particular, 5G-Advanced and 6G networks open the door to non-standard enhancements, especially in private network deployments, where performance optimization can be more flexibly tailored. Recent studies related to network coding have shown great potential in achieving URLLC using mmWave communication \cite{dias2023sliding} as well as significant performance enhancement over ARQ-based repair mechanisms \cite{michel2022flec,esfahanizadeh2024benefits}. In network-sliced environments, network coding enables customized reliability strategies that are well-aligned with application-specific requirements. Similarly, potential standardization opportunities for coding schemes and how they can be integrated into the 5G protocol stack were discussed in \cite{landon2024enhancing}, further supporting their relevance for next-generation network design.

We extend the work in \cite{esfahanizadeh2024benefits,landon2024enhancing} to address expected 6G use cases and showcase that the network coding-based erasure correction mechanism outperforms the multi-layer reliability mechanisms defined in the current standards. Specifically, we show how implementing network coding at a higher layer of the protocol stack could reduce the service time and help meet the stringent requirements of HRLLC, in contrast to the feedback-based HARQ and ARQ mechanisms used in 5G. We elaborate on the theoretical characterizations of our approach and the current standards, and evaluate them in a simulated environment. We showcase various impacts of the network coding implementation in a practical network, demonstrating its ability to support diverse applications with high reliability and low latency over shared infrastructure. Particularly, we show that the in-order delivery delay of packets in practical scenarios can be reduced by half while maintaining comparable throughput for terrestrial networks. We also extend our analysis to scenarios with longer round trip times and error probabilities, representative of non-terrestrial networks. In these settings, network coding continues to deliver performance gains in both delay reduction and throughput improvement. These findings also present a strong case for incorporating advanced techniques such as network coding into future 6G standards. More broadly, our results highlight the importance of designing standards with the flexibility to incorporate non-standardized technologies, enabling adaptation to diverse application requirements. Particularly, there are the following major contributions in this work.

\begin{itemize}
    \item We propose the introduction of a network coding layer above the RLC and MAC layer in the 5G protocol stack to handle erasures.
    \item We propose to bypass the ARQ and HARQ error corrections and use network coding as the reliability assurance mechanism.
    \item We characterize the ARQ/HARQ cross-layer reliability mechanism and compare it with our proposed network coding based approach.
    \item We extend the studies in \cite{esfahanizadeh2024benefits} to include a cross layer reliability mechanism and add additional realistic network environments such as varying RTT, low RTT and high RTT to represent different application scenarios. 
    \item We present simulation studies that directly compare 5G baseline to our proposed approach and showcase the benefits of network coding over the existing reliability mechanisms.
    \item We present a dedicated session on how and why the 6G standardization efforts should consider network coding-enabled erasure correction mechanisms to meet the IMT proposed requirements.
\end{itemize}

The rest of the paper is organized as follows. Section \ref{sec:system_model} presents the protocol stack and describes the proposed system, protocols, and key parameters. Section \ref{sec:theory} focuses on the characterization of our proposed network coding implementation and 5G baseline systems. It is followed by the verification of these theoretical characterizations through simulations in section \ref{sec:results}. We further expand on these results and present how these can impact future networks and 6G standardization efforts in section \ref{sec:discussions} before concluding in section \ref{sec:conclusion}. 

\section{Literature Review}
\label{sec:SOTA}
% Paragraph from intro - Additional works in wireless, QUIC (FlEC, etc), cross layer reliability mechanisms, protocol stack integrations(NC/TCP, SNOB, Yunus, and more)

Reliability mechanisms have been explored in each layer of the network protocol stack. Works such as \cite{barakat2002bandwidth} and \cite{chockalingam1999wireless} analyze the tradeoffs that emerge when link-layer FEC/ARQ is used beneath TCP, finding that often separation of repair and congestion signals results in lost information about channel conditions. In 5G and beyond systems, cross-layer reliability mechanisms are explored as a possibility to achieve URLLC over wireless \cite{ibrahim2025urllc6genabledindustry,karnam2025reliability} and in the context of Industry 4.0 and IoT applications \cite{ramly2021cross,martalo2024cross}. Other works explore coordinating the two processes by bringing erasure correction to the transport layer. QUIC-FEC integrates FEC schemes (XOR, Reed Solomon, and convolutional RLC) with Google's transport protocol QUIC \cite{michel2019quic}. This was extended to use network coding in \cite{michel2022flec}.

Network coding (NC) has also been proposed to smooth loss bursts and avoid head-of-line blocking at the transport layer, via implementations such as TCP/NC \cite{sundararajan2011network}. TCP/NC introduces a coding sublayer beneath TCP to deliver degrees of freedom rather than raw packets, improving throughput over lossy paths without changing the congestion-control API. Coded TCP (CTCP) goes further by adapting the congestion response to coded losses and wireless conditions \cite{kim2012network}. These systems demonstrate that coding near the transport can reduce retransmission delays and variance in delivery time; however, they typically operate at the transport layer agnostic to the configurations in lower layers. Our work complements this investigation by examining the impact of network coding when link-layer reliability mechanisms are turned off.

In industrial 5G NR settings, \cite{paris2020addressing} implements NC across multiple gNode   Bs (gNBs) so that the UE can recover from erasures without per-link feedback, by each gNB sending a copy of the Protcol Data Unit (PDU) at the Packet Data Convergence Protocol (PDCP) layer. In heterogeneous multi-hop backhaul scenarios, adaptive-causal random linear network coding (AC-RLNC) is used as a network service, demonstrating that controller-driven placement of coding/decoding functions can generalize across mesh topologies and traffic classes \cite{cohen2021bringing}. For eMBB video, \cite{vukobratovic2018random} discusses RLNC above or within 3GPP layers to trade redundancy for reduced stall and faster in-order delivery. Furthermore, multiple discussions in the 3GPP RAN meetings over the years included network coding as a potential approach to achieve superior performances in XR applications, Integrated Access and Backhaul (IAB) and as a MAC layer outer-coding scheme over the years \cite{RP212396,RP193220,RP232425}. Although largely architectural, its exploration of different insertion points for NC foreshadows our exploration of NC at the MAC layer.

Across these threads, consistent gaps emerge: feedback-based HARQ/ARQ imposes RTT-sized service-time penalties that are incompatible with HRLLC at realistic BLERs; and existing NC integrations stop short of explicitly bypassing cellular HARQ/ARQ to quantify end-to-end service-time impacts in sliced deployments. Our work directly addresses these by (1) characterizing ARQ/HARQ vs.\ NC delay analytically, and (2) validating in a sliced simulator that NC reduces in-order delivery delay and also lowers network resource usage for coexisting slices.

\section{System Model} \label{sec:system_model}

\begin{figure*}[!t]
  \centering
  \includegraphics[width=0.95\linewidth]{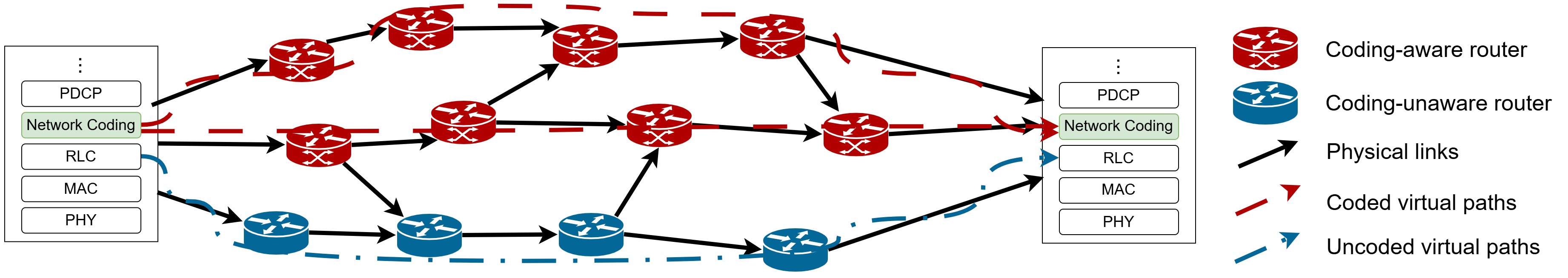}
  \caption{Introduction of new virtual network coding layer in the 5G protocol stack. This virtual network coding interception can be performed above RLC layer for including forward erasure correction capabilities. Packets that goes through this network coding layer will be sent through the virtual paths that are coding aware. The non-coded packets can skip this virtual coding layer and transmit over RLC layers.}
  \label{fig_virtual_path1}
\end{figure*}

We consider a heterogeneous wireless network environment that serves a diverse set of applications over shared infrastructure. In this setting, multiple virtual paths can be defined within a sliced network, each connecting source and destination nodes via distinct physical links. These paths may consist of different channel characteristics, including RTT and frame error probability in the physical layer. In a heterogeneous 6G environment, these paths may have different radio access technologies, including cellular, Wi-Fi, mmWave, or satellite communication paradigms. %Figure ~\ref{fig_virtual_path} shows a scenario with multiple virtual paths over the network and different paths may have different capabilities and characteristics. The paths that include coding-aware routers can use our network coding approach while routers that are not aware of coding schemes can co-exist in the network and be part of uncoded virtual paths. These virtual paths may be assigned to slices based on the requirements.
Current standards of the 5G protocol stack employ cross-layer reliability mechanisms, namely HARQ in the MAC layer and ARQ in the RLC layer, to ensure reliable transmission. In contrast, our proposed architecture introduces a network coding implementation above RLC layer and disables these standard retransmission processes. %As shown in Figure ~\ref{fig_NClayer}, the proposed system inserts the network coding module between the RLC and PDCP layers. 
Figure ~\ref{fig_virtual_path1} shows the introduction of the virtual network coding layer between the PDCP and RLC layer. The coded packets flow through the virtual paths that are coding aware while the uncoded packets may skip the virtual coding layer and connect over RLC as in the traditional 5G systems.
This placement allows coding to operate over end-to-end flows without interference from lower-layer retransmissions. This additional layer does not require any changes in the PDCP or RLC layer as it takes in the northbound PDCP Service Data Units (SDU) (that are passed on as RLC southbound SDUs) and perform network coding before passing the same structure of SDUs to RLC layer. In fact, this network coding activities can also be considered as a sublayer activity for RLC layer. No standardized procedures at RLC or PDCP layer require to be removed for the inclusion of network coding-based reliability mechanism.

% \begin{figure}[!t]
%   \centering
%   \includegraphics[width=0.9\linewidth]{Figures/Final/JSAC_1.jpg}
%   \caption{The network can have multiple virtual paths from a source to destination based on the physical links between the nodes. These virtual paths are assigned to each slice based on the user requirements. Based on the capability of the intermediate nodes to perform coding, some of these virtual paths can employ the network coding scheme we propose in this work.In this figure, two virtual paths are compatible with the coding function while the last virtual path includes coding-unaware routers and thus not capable of implementing coding.}
%   \label{fig_virtual_path}
% \end{figure}

% \begin{figure}[!t]
%   \centering
%   \includegraphics[width=0.9\linewidth]{Figures/Final/JSAC_2.jpg}
%   \caption{The network can have multiple virtual paths from a source to destination based on the physical links between the nodes. These virtual paths are assigned to each slice based on the user requirements. Based on the capability of the intermediate nodes to perform coding, some of these virtual paths can employ the network coding scheme we propose in this work. In this figure, three virtual paths are compatible with the coding function while two paths that include coding-unaware routers are not capable of implementing coding.}
%   \label{fig_virtual_path}
% \end{figure}

For the MAC layer, we limit the HARQ transmissions to 1 and employ Unacknowledged Mode (UM) in RLC layer, to avoid feedback-based ARQ retransmissions. This setting allows the lower layer frames to be passed to the virtual network coding layer without significant delays and use network coding as the reliability mechanism. These architectural changes needs to be performed at both ends of the communication systems. In a Wireless Access Backhaul scenario, the intermediate nodes with full stack capacity can also perform network coding (We limit the discussion on intermediate nodes perform coding to a minimum for brevity and focus on end-to-end coding in this paper). Prior work has explored network coding at various layers, including the IP layer in WiMAX systems \cite{WiMAX_study}, the transport layer in a mmWave testbed \cite{dias2023sliding}, and the MAC or higher layers in 5G systems \cite{landon2024enhancing}. Our work builds on these efforts to support dynamic and application-aware reliability in sliced 6G environments. This proposed approach integrates network coding as a reliability mechanism with minimal changes to the existing 5G protocol stack.

%We assume that each virtual path in this network has the same capacity but a round trip time (RTT) and a probability of error in the data unit transferred taken from a uniform distribution.

% \begin{figure}[!t]
%   \centering
%   \includegraphics[width=0.4\linewidth]{Figures/Final/5GNCLayer.jpg}
%   \caption{Introduction of new virtual network coding layer in the 5G protocol stack. This virtual network coding interception can be performed above RLC layer for including forward erasure correction capabilities.}%\ale{We replace/update an existing layer or add a new one? It is not clear from the figure. Make clear which layer we replace/update in the proposed approach}}
%   \label{fig_NClayer}
% \end{figure}

In a shared network infrastructure, different applications impose diverse performance requirements, particularly in terms of reliability and latency. To address these, we compare the performance of two reliability mechanisms: the state-of-the-art protocol based on ARQ and HARQ, and our proposed NC-aware protocol, which replaces retransmissions with coded combinations of packets. In the current 5G standard, both HARQ and ARQ are used to address packet losses. HARQ is implemented in the MAC layer and operates at the level of code blocks or code block groups, using LDPC codes with up to four predefined redundancy versions. According to 3GPP specifications, if a packet is not successfully decoded after four HARQ transmissions, recovery is delegated to the ARQ mechanism at the RLC layer, which retransmits the full packet. A link failure is declared if no acknowledgment is received after a total of $maxRetxThreshold$ transmission attempts across both HARQ and ARQ layers.

In our proposed approach, we replace the ARQ and HARQ retransmissions in the 5G protocols with an RLNC block coding scheme. In this scheme, additional repair packets (or frames) are proactively generated and transmitted alongside the original data packets to compensate for potential erasures. The original packets coded together as a block is called a generation and the number of original packets is called the generation size, $k_j$. The coding operations are performed in Galois Field (for most practical cases, $GF(2^8)$ is enough) and the size of a coded packet will be same as the largest packet in the generation. Frames with errors in the MAC or RLC layers are treated as erasures and are recovered using the network coding-based repair packets, without the need for feedback-based retransmissions. To enable this functionality, we introduce a virtual NC layer in the protocol stack, just above the RLC layer. It can be added above the RLC layer without modifying higher layers, which remain agnostic to the coding. A similar architecture was successfully demonstrated in prior experiments with the WiMAX protocol stack \cite{WiMAX_study} where network coding was inserted to enhance reliability. This same design principle can be extended to the 5G stack.

We characterize the performance of these protocols for a given slice and analyze how different slicing configurations meet application-specific requirements. A slice is defined by the number of virtual paths allocated to a particular application. We first define the key performance metrics of an application that acquires the $j$-th slice, $j\in\{1,\dots,J\}$. The performance metrics depend on resources that are allocated to the application, i.e., $\mathcal{P}_j$, as well as the deployed communication solution by the application, e.g., RLNC or ARQ and HARQ.
\begin{itemize}
    \item \textbf{(In-Order) Delivery Delay:} Number of time slots it takes for an information packet to be delivered (in-order) at the destination, denoted with {$D(\mathcal{P}_j)$} and {$I(\mathcal{P}_j)$}, respectively.
    \item \textbf{Goodput:} Number of information packets that are delivered per time slot, denoted with {$G(\mathcal{P}_j)$}.
    % \item \textbf{Completion Time:} Number of time slots it takes for an application to successfully deliver $\nu$ information packets, denoted with {$T_\nu(\mathcal{P}_j)$}.
\end{itemize}

To satisfy the diverse requirements of each application, the network must be sliced in a way that aligns resource allocation with performance objectives. Specifically, we aim to ensure that $E[G(\mathcal{P}_j)] \geq \bm{G}_j$ and $E[D(\mathcal{P}_j)] \leq \bm{D}_j$, where $\bm{G}_j$ represents the minimum required \textit{goodput} and $\bm{D}_j$ denotes the maximum acceptable \textit{delivery delay} for the $j$-th slice. This guarantees that the application's quality of service (QoS) remains uncompromised. In Section~\ref{sec:theory}, we present a theoretical framework for evaluating application performance under both the traditional and proposed reliability approaches.

\section{Characterization}\label{sec:theory}

% We may have to characterize HARQ theoretically.
% And extend the completion time.
% HARQ model

In this section, we formalize the performance characteristics of a communication protocol operating under specific slice conditions, including bandwidth, RTT, and packet error probability. Our proposed solution is based on RLNC \cite{ho2006random,toemoeskoezi2014delay,cloud2015coded,9174386,9017940,domanovitz2020streaming,waxman2025blank} and we compare it against a combination of ARQ and HARQ as defined in the 5G protocol stack. We consider a communication system that is time-slotted and, for conceptual clarity, assume a fixed round-trip time ($RTT$) and average packet erasure probability denoted by $\overline{\mathcal{P}_j}$ for the $j$-th slice. This error probability reflects the likelihood of packet loss due to decoding failure or transmission errors over the underlying physical channel. 

\subsection{RLNC-based Communication Protocol}\label{RLNC_C1}

This communication solution is an adaptive RLNC scheme. Consider the $j$-th application operating over an allocated slice $\mathcal{P}_j$, $j\in\{1,\dots,J\}$. For this study, an adaptive block RLNC approach is considered \cite{ho2006random,esfahanizadeh2024benefits}. In this approach, a \textit{generation} is defined as a sequence of original packets that are coded together using randomly generated coefficients; the generation size is denoted with $k_j\gg |\mathcal{P}_j|$. Following the principles of RLNC, each coded generation includes $k_j\gamma_j^1$ coded packets. The initial coding parameter $\gamma_j^1$ is called the FEC rate, and can be decided to optimize the performance of the network. A reasonable choice is proportional to $1/(1-\overline{\mathcal{P}_j})$, to balance resource usage and erasure correction capability. A larger $\gamma_j^1$ will increase the erasure correction possibility but may use more resources. The sender sequentially transmits the coded generations, starting with the first generation. For each generation, it transmits $k_j\gamma_j^1$ coded packets through the links in $\mathcal{P}_j$. Upon receiving feedback of the last-sent encoded packet of a generation, the sender can verify whether the receiver requires any additional coded packets, missing degrees of freedom denoted by ${m}_j$, to decode the generation. It then transmits $\gamma_j^2 {m}_j$ coded packets in the next time slot. The coding parameter $\gamma_j^2$ is called the feedback-based (FB) rate, and is set such that the probability that the receiver fails to decode the generation after the second round is almost zero. These feedback based transmissions are only required if the initial FEC packets were not enough to correct erasures. By selecting appropriate $\gamma_j^1$, the sender can handle latency-goodput tradeoffs. %If the sender has the option to select among the paths, its strategy is a random selection.

We first characterize the random variable ${m}_j$. The probability of ${m}_j = 0$ is equal to having equal or less than $\lceil(\gamma_j^1-1)k_j\rceil$ failed transmissions in the first trial of a generation. Similarly, the probability of ${m}_j = m$, $0<m\leq k_j$, is equal to having  $\lceil(\gamma_j^1-1)k_j\rceil + m$ failed transmissions in the first trial. 

\begin{lemma}\label{lemma:prob_missing_NC}
Distribution of the random variable $m_j$ is approximated as follows,
    \begin{equation}
    \label{dist_m}
    P[{m_j}=m]\approx\begin{cases}
        \sum_{f=0}^{\left\lceil\left(\gamma_j^1-1\right)k_j\right\rceil}\frac{\lambda^{f}e^{-\lambda}}{f!}&m=0,\\\vspace{-0.2cm}\\
        \frac{\lambda^{\left\lceil\left(\gamma_j^1-1\right)k_j\right\rceil + m}e^{-\lambda}}{\left(\left\lceil\left(\gamma_j^1-1\right)k_j\right\rceil+m\right)!}&m\in\{1,\dots,k_j\},\\\vspace{-0.2cm}\\
        0&\text{otherwise},
    \end{cases}
\end{equation}
where $$\lambda=k_j\gamma_j^1\overline{\mathcal{P}_j}.$$ 

\end{lemma}
\begin{proof}
    The number of failures in a trial can be modeled as the sum of independent non-identical Bernoulli random variables. If the erasure probabilities (Bernoulli parameters) are close to zeros, this distribution can be approximated by a Poisson distribution with its parameter being the summation of the parameters of the Bernoulli random variables \cite{10.1214/aoms/1177705799}, i.e., $$\lambda=\frac{k_j\gamma_j^1}{|\mathcal{P}_j|}\sum_{p_i\in\mathcal{P}_j} p_i=k_j\gamma_j^1\overline{\mathcal{P}_j}.\vspace{-0.5cm}$$
\end{proof}
\begin{corollary}
    The average delivery delay for RLNC is,
    \begin{equation}
        \label{eq:AVG_D_RLNC}
        \begin{split}
        E[D_R(\mathcal{P}_j)] \approx & \left(\frac{RTT}{2}+\left\lceil{\frac{k_j\gamma_j^1}{|\mathcal{P}_j|}}\right\rceil\right)P[{m_j}=0]\\
        & \hspace{-2 cm}+\left(\frac{3RTT}{2}+\left\lceil{\frac{k_j\gamma_j^1}{|\mathcal{P}_j|}}\right\rceil+ \left\lceil{\frac{m_j\gamma_j^2}{|\mathcal{P}_j|}}\right\rceil\right)(1-P[{m_j}=0]).
        \end{split}
\end{equation}
\end{corollary}
\begin{proof}
    The delivery delay of a packet depends on how many trials it takes for the generation of that packet to be decoded at the receiver. Thus, the distribution of delivery delay for the presented RLNC solution is,
\begin{equation*}
    D_R(\mathcal{P}_j)\approx
    \begin{cases}
        \frac{RTT}{2}+\left\lceil{\frac{k_j\gamma_j^1}{|\mathcal{P}_j|}}\right\rceil,& {m_j}=0,\\\vspace{-0.2cm} \\
        \frac{3 RTT}{2}+\left\lceil{\frac{k_j\gamma_j^1}{|\mathcal{P}_j|}}\right\rceil+ \left\lceil{\frac{m_j\gamma_j^2}{|\mathcal{P}_j|}}\right\rceil,& {m_j}>0.
    \end{cases}
\end{equation*}
Therefore, the average delivery delay for RLNC can be obtained as in (\ref{eq:AVG_D_RLNC}).\footnote{Similarly, one can identify the second moment and variance of delivery delay, which is not presented in this paper for brevity. }
\end{proof}

\begin{corollary}
The average goodput for RLNC is,
\begin{equation}
\label{eq:AVG_G_RLNC}
    \begin{split}
        E[G_R(\mathcal{P}_j)]&=E\left[ \frac{k_j}{k_j\gamma_j^1+{m_j}\gamma_j^2}|\mathcal{P}_j|\right]=\\&\sum_{m=0}^{k_j} \frac{k_j}{k_j\gamma_j^1+m\gamma_j^2}P[{m_j}=m]|\mathcal{P}_j|.
    \end{split}
\end{equation}
\end{corollary}

\subsection{ARQ/HARQ Mechanisms in 5G Systems}\label{5G_C2}
In this setting, we characterize the reliability mechanism employed in the current 5G NR systems, which serve as our baseline protocol. Under current standards, the reliability mechanisms in 5G are based on HARQ in the MAC layer and ARQ in the RLC layer. At the MAC layer, HARQ attempts to recover from frame-level decoding failures, i.e., when a transport block cannot be decoded correctly after physical layer demodulation. To correct such errors, HARQ uses incremental redundancy with LDPC codes. The initial transmission of HARQ includes some parity bits that are capable of correcting a limited number of bit errors. However, if the parity bits cannot recover the original transport block, additional LDPC codewords are sent in the next iterations. After a specific number of such iterations, defined by $maxHARQTx$, if the errors are still not corrected, the RLC layer retransmissions start. This will initiate a new round of HARQ retransmissions at MAC layer. If the total number of attempts exceeds a predefined threshold, $maxRetxThreshold$, the system declares a transmission failure. Typical HARQ and ARQ transmission limits are 4 and 8, respectively \cite{ding2021optimized,moothedath2025delay}.

\subsubsection{Delivery delay}
To characterize this setting and derive the theoretical performance analysis, we provide a separate characterization of ARQ and HARQ. Then we will combine it to represent the 5G system. For ARQ, the approach is simple: Whenever a packet loss is reported to the RLC layer (i.e., after $maxHARQTx$ HARQ transmissions), it is retransmitted completely and independently. Assuming that the channel conditions remain the same till the retransmission, the retransmission has the same probability of success. According to the definition, \textit{delivery delay} of a packet for the $j$-th application $D_A(\mathcal{P}_j)$ is the difference between the time the packet is transmitted for the first time through its allocated slice and the time it is successfully received. We assume the transmission time of a packet (including the first transmission and the possible re-transmissions) is negligible compared to $RTT$, and do not consider it in the following deviations. Thus, 
\begin{equation*}
    D_A(\mathcal{P}_j)\in\left\{\frac{RTT}{2},\frac{RTT}{2}+RTT,\frac{RTT}{2}+2RTT\dots\right\}.\vspace{-0.1cm}
\end{equation*}

\begin{lemma} \label{lemma:dist_D_MP_SR_ARQ}
Distribution of delivery delay for ARQ is,
    \begin{equation}
        \label{eq:PMF_D_ARQ}P\left[D_A(\mathcal{P}_j)=\frac{RTT}{2}+kRTT\right]=
            \overline{\mathcal{P}_j}^k\left(1-\overline{\mathcal{P}_j}\right),
    \end{equation}
    when $k\in\{0,1,\dots\}$, and zero otherwise. Here, $\overline{\mathcal{P}_j}=\left(\sum_{p_i\in\mathcal{P}_j} p_i\right)/{|\mathcal{P}_j|}$ is the average erasure probability of the slice. %Besides,
\end{lemma}

\begin{proof}
    Every packet has an initial delay of $\frac{RTT}{2}$ to travel from transmitter to receiver. Each additional retransmission incurs an additional RTT of delay. The probability that a packet will have a delivery delay of $\frac{RTT}{2}+kRTT$ is the product of the probability of $k$ failed transmissions (given by $\overline{\mathcal{P}_j}^k$) and one successful transmission (given by $\left(1 - \overline{\mathcal{P}_j} \right )$).
\end{proof}

HARQ is an enhanced version of SR-ARQ that combines the retransmission system with forward error correction. Past transmissions are saved and combined to allow a complete decoding from multiple incomplete fragments, if necessary, thus improving the probability of reception with each retransmission.

The definition and potential values of the \textit{delivery delay} of a packet for the $j$-th application $D_H(\mathcal{P}_j)$ is unchanged from that of SR-ARQ,
\begin{equation*}
    D_H(\mathcal{P}_j)\in\left\{\frac{RTT}{2},\frac{RTT}{2}+RTT,\frac{RTT}{2}+2RTT\dots\right\}.\vspace{-0.1cm}
\end{equation*}

The probability distribution, however, must take into account the reduced probability of error with each additional retransmission, due to the HARQ mechanism. Thus, before defining the probability distribution of the delivery delay, we establish an approximation for the effective Signal to Noise Ratio (SNR) of retransmission. The effective SNR of a message transmitted $N$ times with HARQ is approximately $N$ times the SNR of that message transmitted once, and the probability of error for a given message is a function of that effective SNR, as explained in detail in \cite{landon2024enhancing}.

Assuming that SNR is approximately the same across retransmissions of a single message, we can use this approximation to define the distribution of the delivery delay for HARQ.

\begin{lemma} \label{lemma:dist_D_MP_HARQ}
Distribution of delivery delay for HARQ is, for $\sigma$ denoting SNR in linear units,
    \begin{equation}
        \label{eq:PMF_D_HARQ}
            \begin{array}{@{}l@{}}
            P\left[D_H(\mathcal{P}_j)=\dfrac{RTT}{2}+kRTT\right] =\\
            \hfill
            \displaystyle 
            \begin{cases}
            1 - \overline{\mathcal{P}_j}(\sigma) & \text{when } k = 0\\[1.2ex]
            \left(1 - \overline{\mathcal{P}_j}((k+1)\sigma)\right)
            \prod_{i=1}^{k} \overline{\mathcal{P}_j}(i\sigma) & \text{when } k > 0
            \end{cases}
            \end{array}
    \end{equation}
    or can be represented as:
    %\[
  \begin{multline*}
       P\left[D_H(\mathcal{P}_j) = \frac{RTT}{2} + k \cdot RTT\right]\\ =
        \left(1 - \overline{\mathcal{P}_j}((k+1)\sigma)\right)
        \cdot \prod_{i=1}^{k} \overline{\mathcal{P}_j}(i\sigma)].
    \end{multline*}  
    %\]
    We define the empty product as \(\prod_{i=1}^{0} (\cdot) = 1\) to simplify the expression for $k = 0$.
\end{lemma}

\begin{proof}
    As in Lemma \ref{lemma:dist_D_MP_SR_ARQ}, the potential delivery delays are some $(k+1/2)$ multiple of the RTT. However, due to the LDPC error correction implemented in HARQ, the probability of successful transmission depends on SNR and increases as a function of the product of SNR and $k$ as described in \cite{landon2024enhancing}. 
\end{proof}

Now, combining both HARQ and ARQ approaches, our baseline 5G systems can be characterized as follows. For first $maxHARQTx$ transmissions, we follow the distribution exactly as defined for the HARQ scenario in Lemma \ref{lemma:dist_D_MP_HARQ}. After the maximum number of transmissions, the ARQ retransmission mechanism kicks in, initiating another set of HARQ transmissions. However, this time the SNRs once again start from $\sigma$ and get incremented. This scenario can be represented as:

    \begin{equation}
        \label{eq:PMF_D_Baseline}
            \begin{array}{@{}l@{}}
            P\left[D_{AH}(\mathcal{P}_j)=\dfrac{RTT}{2}+kRTT\right] =\\
            \hfill
            \displaystyle 
            \left(1 - \overline{\mathcal{P}_j}((k_r+1)\sigma)\right)\cdot P(H_e)^r\cdot 
            \prod_{i=1}^{k_r} \overline{\mathcal{P}_j}(i\sigma),
            \end{array}           
    \end{equation}
where 
$P(H_e)$ is the probability of a complete round of HARQ transmissions failing to succeed, \[ P(H_e) = \prod_{i=1}^{maxHARQTx} \overline{\mathcal{P}_j}(i\sigma),\] \\
$r$ is the number of HARQ rounds completed before this round, \[ r= \floor{k \div maxHARQTx}, \]\\
and $k_r$ is the number of HARQ transmissions in that particular round, \[k_r = k \mod maxHARQTx.\] 
When $r$ exceeds the maximum number of ARQ transmissions allowed ($maxRetxThreshold$), a connection failure is defined and the probability of a successful reception will be 0 for such a case.

\subsubsection{Goodput} 
We have defined goodput as the number of information packets that are delivered per time slot and denoted by $G_{AH}(\mathcal{P}_j)$. Thus, the goodput per any slot is equal to the number of links that had a successful transmission at that slot. Let's consider the slice $j$ with $|P_j|$ number of individual links in the slice and each link has a probability of error $p_i$. However, with HARQ-ARQ combined reliability mechanism, the probability of error is a function of $\sigma$. Specifically, we must distinguish between the first HARQ transmission and its subsequent retransmissions. Thus, the expected goodput at any particular slot for a slice $j$ with $|P_j|$ individual links and each link having a probability of error $p_i$ can be defined as:
\[
E[G_{AH}(\mathcal{P}_j)]=\sum_{p_i\in\mathcal{P}_j} (1-p_i(k_r\sigma)), \label{eq:AVG_G_ARQ}
\]
where $k_r \in [1,maxHARQTx]$. 

% This definition of goodput is packet-level, not byte-level; a network-coded packet will require an additional few bytes to communicate a seed to generate coding coefficients and other information. In one of our implementations, this header is 3 bytes. The impact of this overhead varies with the average packet size, which depends on the layer at which network coding is implemented. Transport blocks on the MAC layer range up to a million bits while frames on the IP layer generally have a maximum transmission size of 1500 bytes. In each case, the overhead of a network coding header is a fraction of a percent of the packet size, so benefits of packet-level goodput will not be significantly reduced when taking byte-level goodput into account.

It is to be noted that the probability of success in any particular transmission is a function of the instance of HARQ transmission/retransmission. The probability of success increases with $k_r$, but the exact probability is empirical. With the complexities in the setting of our baseline 5G standard system, finding a closed-form solution for either the average delivery delay or the goodput is not our goal. Thus, for our comparisons, even though RLNC-based approach provides closed-form solutions for the performance metrics, we rely on real-time system simulations. Similarly, we rely on simulations for in-order delivery delay results in this work.

\section{Simulation and Results}\label{sec:results}

\subsection{Simulation Settings}
In this section, we expand upon our theoretical characterization of the network with mixed slices. We validate the results empirically by simulating real-time networks while further studying performance criteria and slicing strategies. In particular, we explore our implementations of both NC-aware protocol and the state-of-the-art 5G protocol (as baseline) through SimPy, a discrete-event simulator \cite{simpy}. SimPy excels in simulating real-world networking scenarios thanks to its adeptness in handling asynchronous events, time-dependent behaviors, and custom event scheduling. By adopting a process-based paradigm, SimPy enabled us to effectively simulate data flow within the network and replicate desired client-server transactions.

To simulate the non-coding scenario, we consider the standard 5G protocol stack, with both ARQ and HARQ as the reliability mechanisms. For the coding case, we consider network coding applied above the RLC layer, bypassing any ARQ/HARQ-type retransmissions. The error correction using HARQ retransmissions are empirically demonstrated. In our characterization, we show how the probability of success in a retransmission of an HARQ packet can be approximated by the probability of success at half the noise level.

In order to replicate HARQ's decrease in erasure probability upon subsequent re-transmissions, we ran the MATLAB 5G toolbox simulations. The error correction capability of the 5G HARQ process using incremental redundancy through LDPC codes is replicated in the toolbox \cite{MathWorks2025_NR_HARQ}. We did an empirical analysis of the performance of HARQ using the toolbox for the initial block error rate of 5 to 20\%, aligned with realistic channel conditions for our target use cases. Our empirical analysis for fixed MCS at MCS 7 and MCS 14 targetting a 10\% BLER after first transmission showed that each retransmission reduced the erasure probability by half, matching the theoretical analysis presented in Section \ref{5G_C2}. In fact, the analysis at MCS 7 showed that the erasure probability after a retransmission is higher than half of the initial probability. However, we consider that each retransmission reduces the erasure probability by half for brevity in the SimPy simulations. Based on this analysis, we model the 5G protocol stack performance as follows: for a given channel error probability $p_i$, with each HARQ transmission, the error probability is reduced by half. If the transport block is not successfully received, it is retransmitted as a new packet (ARQ) and undergoes another four rounds of HARQ transmissions, starting again with $p_i$ as the error probability for the first transmission. There are up to 8 such rounds of HARQ-ARQ retransmissions, totaling 32 transmissions to send a particular transport block before a failure is confirmed, as per standard procedures. This approach demonstrates the closest comparison to the existing 5G protocols, where the MAC and RLC layers provide reliability using ARQ and HARQ, and we compare this to our network coding-based approach.

For the network coding case, HARQ and ARQ retransmissions are disabled, but additional coded packets are introduced in each block to compensate for erasures. In our approach, the FEC code rate is set to $\gamma_j^1 = \lceil 1/\mathcal{P}_j \rceil$. However, if any block is not successfully received after the initial transmission of all packets, an additional $2 \lceil 1/\mathcal{P}_j \rceil \times m_j$ packets are sent as feedback-based retransmissions. Assuming that the channel conditions do not change significantly,this ensures that no further retransmissions are required. In the extreme cases where packets are still not received after this partial block retransmission, a link failure is reported. 

For our experiments, we consider a network with a total of $n=20$ links, and two slices ($J=2$), trying to serve two different applications. The first slice is dedicated to the first application, and is allocated $i$ out of the total available links, $i\in\{1,\dots,20\}$. The second slice serves the second application with the remaining resources. The slicing index is defined by the number of links assigned to the first slice. For brevity, we present each scenario considering a single radio access technology, thus we also consider each link has the same capacity. This does not hurt the generality of our system model as paths with more bandwidth can be represented with multiple paths in our model. This accounts for a larger throughput for a slice with larger number of links allotted. The error probability and round-trip time of each link are defined in each experiment. In a practical scenario, it is not necessary that a slice serves only one application. Rather it could serve multiple applications within the same QoS requirements. The QoS Flow Identifier indicates whether the traffic corresponds to URLLC or eMBB slices. Depending on the QoS requirements of each application, a coded slice or an uncoded slice can be used. The packet headers can include additional information like code rate and generation size to facilitate appropriate coding procedures. Further details on the system design perspectives are presented in subsection \ref{system_discussion}.

In all experiments, each application is sending 10,000 packets and the results show the average of 100 iterations of the experiment. We consider multiple scenarios with both RLNC and HARQ as our reliability mechanisms for a highly reliable system that targets 99.99\% of packets to be successfully delivered. The first set of experiments replicates a network that is similar to the 5G NR scenario. For the terrestrial networks, the 3GPP standards limit the number of parallel HARQ processes that a UE can handle at the MAC layer to 16. Thus, any UE following the current standard may have a 16-frame RTT as a maximum in the MAC layer. Considering that the RTT is in the range of 16 slots and an erasure probability of 0.1, the first set of experiments provides a comparison of HARQ performance against an approach where we consider network coding to replace HARQ in the MAC layer directly. A generation size ($k_j$) of 5 is considered for this set of experiments. We present the comparison with fixed RTT and probability of error as well as a more realistic setting in practice where the RTT and erasure rates may not be exact. In this realistic setting, we sample the RTT for each channel from a gaussian distribution and the erasure probability from a Uniform distribution, using the previous fixed constants as our averages.

\subsection{Results and Analysis}

\begin{figure*}
  \centering
  \includegraphics[width=\linewidth]{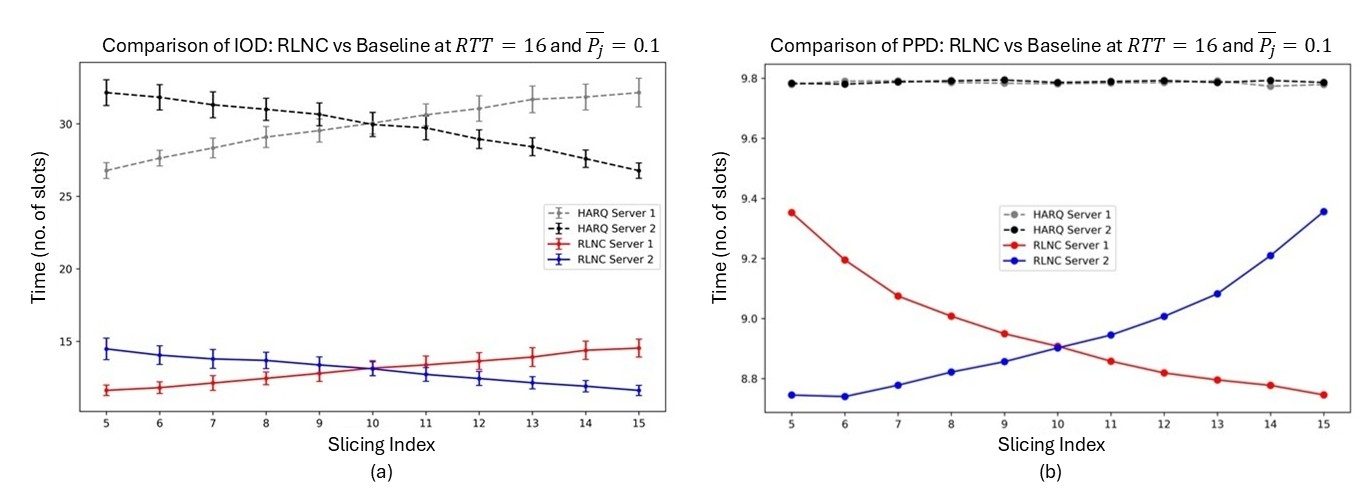}
  \caption{Comparisons of In-order Delivery Delay (IOD) and Per Packet Delivery Delay (PPD) for network coding and 5G baseline with fixed parameters.}
  \label{16Const}
\end{figure*}

Figure \ref{16Const} shows the average delivery delays for both the baseline and network coding systems. It is evident that network coding ensures a much lower in-order delivery delay (IOD). Furthermore, the average per-packet delivery delay (PPD) is also lower compared to the baseline. It can be noticed that the average PPD for our approach is closer to $RTT/2$ since most packets are delivered on the first block transmission itself. This showcases the efficiency of our FEC approach. More interestingly, with larger slices, network coding shows a steady decrease in the average delivery delay, while the baseline systems do not seem to have any benefit from allocating more resources. Furthermore, Figure \ref{16Const}(a) also includes the standard deviation of IOD for both scenarios. It is evident from the figure that network coding ensures more consistent IOD, that enables to provide stronger guarantees on the performance. 

\begin{figure*}
  \centering
  \includegraphics[width=\linewidth]{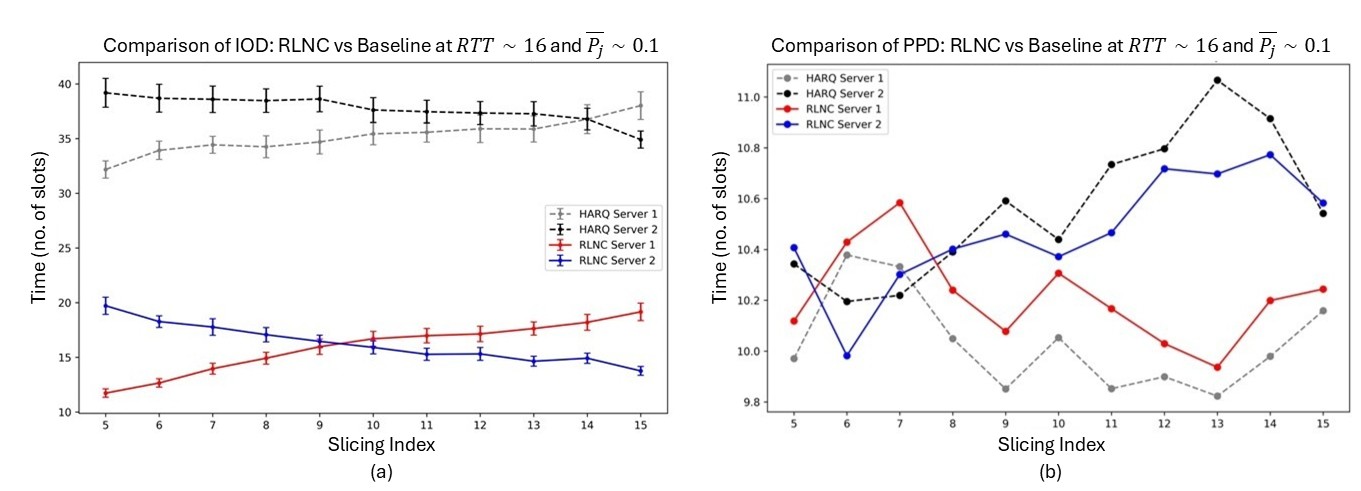}
  \caption{Comparisons of IOD and PPD for network coding and 5G baseline for a varying RTT and error probability.}
  \label{16Delay}
\end{figure*}

Figure \ref{16Delay} shows the delivery delays for the second set of experiments, where the RTT and erasure rates are sampled from a Gaussian distribution. In that case as well, the in-order delay is still significantly lower, but the average delivery delay per packet is comparable for both network coding and 5G baseline, with network coding sometimes having a higher delay. This arises from the fact that the block network coding decodes the packets of a block together and with the low RTT and with flexible error rates, the inherent delay in decoding a block makes the average delay per packet comparable to the traditional coding scheme. However, this issue can be addressed by deploying a sliding window coding approach that allows on-the-fly decoding.

\begin{figure*}
  \centering
  \includegraphics[width=\linewidth]{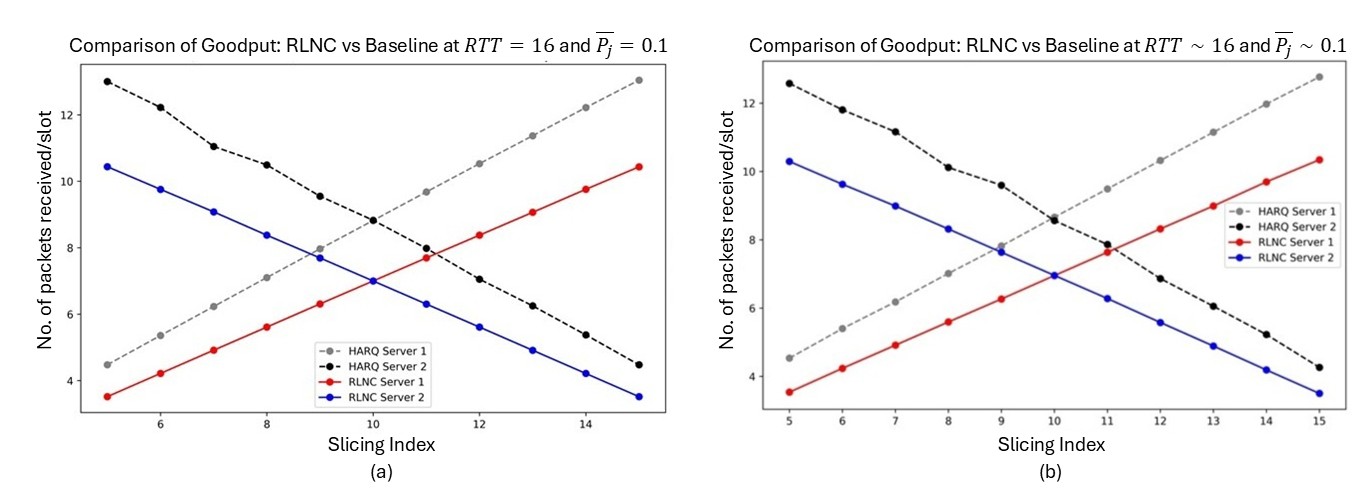}
  \caption{Goodput evaluation of network coding and 5G baseline for low RTT scenarios.}
  \label{16CPGT}
\end{figure*}

Figure \ref{16CPGT} shows the average goodput for these two sets of experiments, where the RTT and error rate are kept fixed or averaged at 16 slots and 10\% respectively. It has to be noted that the goodput in both cases follows the same pattern and is identical, aligning with our theoretical characterization that showed goodput to be dependent only on the average error probability when the links have similar capacity. The goodput of block RLNC is slightly lower compared to the baseline approach, in terms of original packets delivered per slot. However, it has to be considered that the baseline systems have additional coding overhead that arise from the LDPC coding used for HARQ that requires a larger number of bits transmitted per original packets. If we consider the LDPC coderate and convert the goodput to bits per second, network coding may have better goodput than the baseline systems.

\begin{figure*}%[!t]
  \centering
  \includegraphics[width=\linewidth]{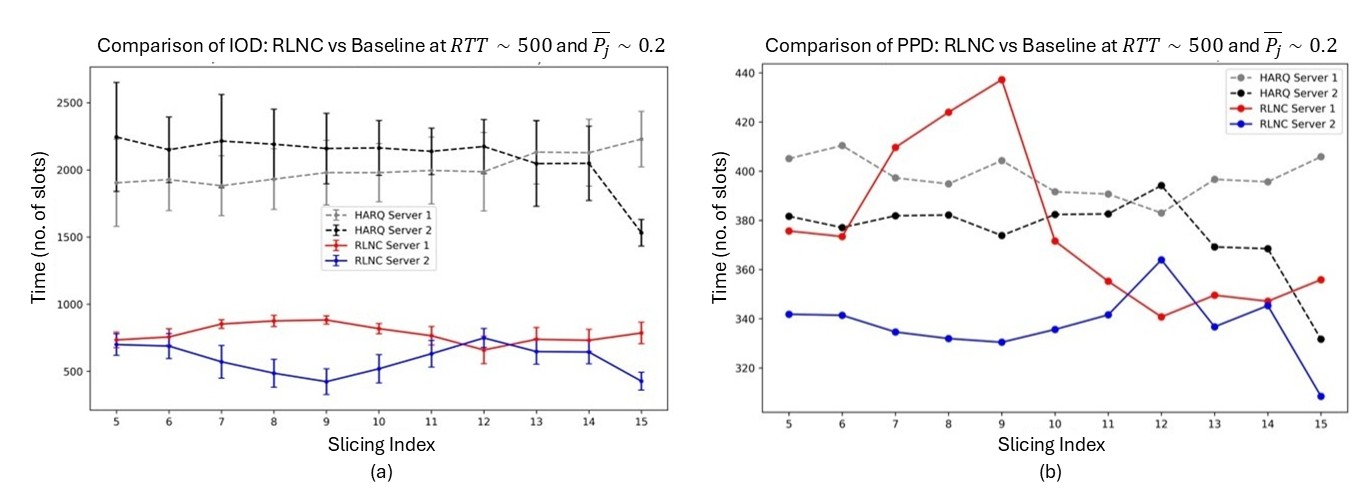}
  \caption{Comparisons of IOD and PPD for network coding and 5G baseline for a larger RTT and higher error probability.}
  \label{500Delay}
\end{figure*}

\begin{figure*}%[!t]
  \centering
  \includegraphics[width=\linewidth]{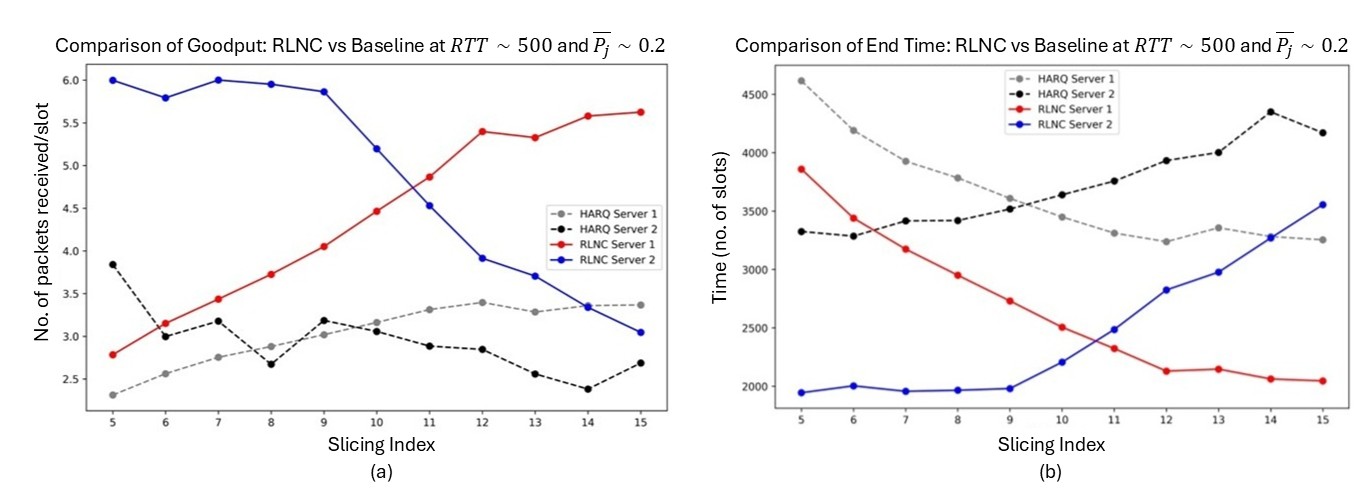}
  \caption{Goodput and completion time evaluation for a large RTT scenario.}
  \label{500CPGT}
\end{figure*}

In the third set of examples, we focus on other scenarios such as non-terrestrial networks or Wi-Fi connected networks, where you may have a higher RTT and error probability. We model this with a high End-to-End delay of 500 slots and an average block error rate of 20\%. We again compare our RLNC approach to the current standards of the 5G protocol stack with RLC and MAC layer error correction mechanisms based on ARQ and HARQ respectively. In our proposed approach, these lower-layer error correction mechanisms are turned off and the network coding layer takes complete responsibility for error correction. Figure \ref{500Delay} shows that the benefits of network coding in delivery delays continue to show as expected in these higher RTTs and Figure \ref{500CPGT} shows that the completion time and goodput also provide better results for the network coding approach compared to the baseline systems. Here the generation size ($k_j$) is 50. Particularly, the completion time results are interesting as they showcases the forward erasure capability of network coding that ensures that when the RTT and erasure rates are higher, in systems like NTN or Wi-Fi, network coding becomes the natural choice for error correction.

\section{Discussion}\label{sec:discussions}

This section elaborates on key findings from our results, emphasizing the advantages of network coding in wireless networks and their potential impact on future wireless standardization efforts.

\subsection{Key Takeaways from the Results}
Some of the key observations from our results that highlight the benefits of network coding are presented  below.

\textbf{Highlight 1: }\textit{A coded slice significantly reduces (in-order) delivery delay and ensures efficient resource utilization for performance improvement.}

Our simulations demonstrate that the network coding approach ensures a reduction in both the average PPD and IOD compared to an uncoded slice. Furthermore, our delay characterization and simulation results indicate that, in the uncoded scenario, the delivery delay of an application depends solely on the average error probability of the available resources. Consequently, even if additional resources are allocated to a slice requiring ultra-low latency, the delivery delay will remain unchanged if the average error probability remains the same. However, with network coding, the delivery delay is influenced by the available resources, leading to improved delay performance as more resources are provided. Notably, the performance gains are not due to increased resources or slicing per se. With the same resources, a coded slice exhibits lower latency than an uncoded slice, indicating that the improvement arises from network coding and would persist even if the network operated as a single slice. Therefore, for achieving low-latency performance in a sliced network, using a coded slice is the more effective approach. Furthermore, our results include standard deviation of IOD to highlight that the variation in IOD is  significantly lower for coded slices, making it more suitable to provide performance guarantees. In fact, with more finely tuned coding parameters, the IOD and its standard deviations can be theoretically calculated to provide a service level agreement with more accuracy compared to the traditional setting.

\textbf{Highlight 2: }\textit{A coded slice satisfies hyper-reliable low latency requirements with fewer network resources, which is not possible with an uncoded setting. Thus, in a slicing scenario, it is better to use coding-enabled resources for an HRLLC slice and allow uncoded resources for an eMBB slice.}

Our analysis shows that a coded slice can achieve HRLLC requirements with fewer network resources compared to an uncoded approach. In an uncoded setting, even if additional resources are allocated, the delivery delay remains constrained by the average error probability of the available channels. However, network coding compensates for the impact of errors, enabling a more efficient use of resources while meeting stringent latency and reliability demands. Consequently, in a slicing scenario, it is advantageous to allocate coding-enabled resources to the HRLLC slice to optimize performance. Meanwhile, enhanced mobile broadband (eMBB) slices, which prioritize throughput over ultra-low latency, can effectively operate with uncoded resources, ensuring a balanced and efficient resource allocation strategy across different network services. This also means that in a network environment with both coding-capable devices and devices without that capability, it is beneficial to allocate devices that support network coding to the HRLLC slice, while assigning those that do not to other applications.

\textbf{Highlight 3: }\textit{For networks with higher RTTs and/or error probabilities, the benefits of network coding are even greater; completion time and goodput also improve in such scenarios.}

When comparing the 16-slot RTT scenario for completion time and goodput metrics, network coding does not provide significant improvements. This is because, in a non-coding setting, reliability can be achieved through retransmissions within the RTT window, allowing lost packets to be recovered efficiently without additional coding overhead. In contrast, network coding introduces some level of redundancy, which in shorter RTT scenarios may not always translate to a performance gain (there are adaptive network coding mechanisms such as \cite{cohen2020adaptive} that can be used to address this issue, however, our analysis here is based on a block coding approach). However, in a long RTT scenario, the inefficiency of waiting for feedback before retransmitting lost packets becomes more evident. Each round-trip delay adds significant latency, slowing down the completion time and reducing goodput. In such cases, network coding mitigates this inefficiency by proactively encoding packets, reducing the dependency on feedback-based retransmissions. This leads to faster completion times and higher goodput, making network coding particularly advantageous in high-latency environments. 

\subsection{System Design Perspectives and Future Works}\label{system_discussion}
The network coding based FEC scheme as a reliability mechanism not only proves the superior performance in delay and reliability but also comes with a low computational and communication overhead. This subsection expands on the discussion of some practical considerations for an efficient implementation of our proposed approach as well as possibilities for improvements.

From a communication perspective, a network-coded packet will require an additional few bytes to communicate a seed to generate coding coefficients and other information. In one of our implementations, this header is 3 bytes. The impact of this overhead varies with the average packet size, which depends on the layer at which network coding is implemented. Transport blocks on the MAC layer range up to a million bits while frames on the IP layer generally have a maximum transmission size of 1500 bytes.  While we do not assume the transport blocks are of same size, we expect that on average, the overhead of a network coding header is a fraction of a percent of the block size, so benefits of packet-level goodput will not be significantly reduced when taking byte-level goodput into account. Furthermore, additional signalling to notify the end users about network coding implementation requires either a flag in the packet header or a system level indication during the connection establishment phase using control signals. Similarly, we expect the UEs and gNBs to have a memory buffer to store a generation till it is either successfully acknowledged or dropped, which in our case was less than 1 MB. This buffer depends on the generation size and the round trip time and is comparable to the requirements of HARQ-based reliability mechanisms. Detailed discussion of the engineering aspects of network coding implementationss including queueing analysis and memory constraints can be found in \cite{medard2025network}.

The computational complexity of network coding operations is minimal compared to the complex decoding process of HARQ. The encoding and decoding process includes matrix multiplications on the second and third order of the generation size, where the complexity increases with the generation size \cite{heide2008cautious}. The generation size of our simulations is 50, ie, 50 packets are coded together to get additional coded packets. A detailed discussion on the generation size from a delay-throughput trade-off perspective is present in \cite{adams2014delay}. 
% Based on the QoS Flow Indicator, each slice can decide its code rate and generation size for that specific PDU session.
Each slice can select a code rate and generation size appropriate to its target parameters. This process can also be made further granular to the level of each PDU session, where code rate and generation sizes can be based off of the QoS flow indicator.
Furthermore, the coding/decoding algorithms do not change depending on the code rate or generation size making it easy to adapt to different QoS requirements. It is recommended to keep the generation size to a small number to reduce delay and jitter. Another approach to reduce jitter is to use a sliding window network coding approach. This provides further improvements in the performance and advanced adaptive schemes based on sliding window \cite{cohen2020adaptive} can improve both the delay and throughput performance. Our future works will explore sliding window network coding implementations.

Another major direction of expanding this work includes comparison of our proposed approach to an adaptive MCS scheme. Currently, we compare our approach to a fixed MCS scenario in the 5G systems. However, many practical applications makes use of an adaptive MCS approach where the MCS values are reduced on the instances of lost packets at a higher MCS value. An adaptive network coding approach that changes the code rate depending on channel quality indices would be equivalent to the adaptive MCS scheme and this comparison is planned as an immediate future work. Similarly, network coding implementations in transport layer have been considered to enhance performance of mmWave scenarios to achieve URLLC performance in mmWave applications \cite{dias2023sliding} and to provide adaptive and flexible QUIC implementations \cite{michel2022flec}. Our approach is complementary to these works and could be compared to analyze the improvements due to lower layer implementation of coding. %It is to be noted that the current approach of MCS based reliability-throughput trade offs have their own benefits and drawbacks. While adaptive MCS scheme ensures that an appropriate modulation order and code rate is chosen based on the channel quality, it requires that the system operates at an MCS value that can guarantee reliable transmissions, at max 10\% BLER after the first transmission. This means that the system operates in a fashion that, for any particular SNR value, an MCS to the left that can guarantee the reliable transmission is chosen. However, this may result in a loss of 3 to 4 dB in the lower MCS values, as shown in Fig. \ref{MCS_1}. However, network coding provides an alternative approach, by operating at an higher MCS with additional NC-enabled reliability mechanism. This provides additional operating points for reliable communications with the minimal coding overhead.

% \begin{figure}%[!t]
%   \centering
%   \includegraphics[width=0.9\columnwidth]{Figures/Final/Figure_HARQ1_optimizedPNG.png}
%   \caption{Evaluation of different MCS operating points against SNR}
%   \label{MCS_1}
% \end{figure}

\subsection{Proposals for 6G Standardization}
The standardization process for 6G has just begun, with the International Telecommunication Union (ITU) releasing the "Framework for IMT-2030 and Beyond" \cite{wp5d2022future} and the "Future Technology Trends" document \cite{wp5d2023m}. The telecommunication industry has also responded with discussions in 3GPP forums, including the first 6G workshop held in March 2025 and the study item for 6G radio is commissioned in the June 2025 plenaries \cite{6GR20}. With usage scenarios such as hyper-reliable low-latency communication and immersive XR/VR applications, ITU-R-2160 highlights the need for advanced technologies that provide high throughput without compromising latency or reliability. Furthermore, techniques that support efficient spectrum utilization, device-to-device communication, and connectivity for a massive number of devices will be crucial to enabling the 6G environment. The application of network coding in such scenarios has already been explored in the literature \cite{pahlevani2014novel, keshtkarjahromi2018device, esfahanizadeh2024benefits} and could become an integral part of 6G standardization efforts. In this section, we focus on some of the recent efforts in 3GPP that include network coding concepts, as well as potential use cases and applications highlighted in ITU-R-2516 where network coding can be beneficial.

Network coding has been discussed, albeit infrequently, within standardization bodies such as 3GPP in the past. Recently, in 2023, a MAC layer outer-coding scheme for reliable single TTI communication was presented in \cite{RP232425}. Intel has been leading efforts on using network coding in Integrated Access and Backhaul \cite{RP193220}, and ZTE has highlighted its potential in achieving the desired performance requirements of XR applications \cite{RP212396}. Similarly, it has also been discussed as a potential technology to address burst loss scenarios in mmWave systems \cite{NGA_WP,biyikoglu2025modeling}. In this paper, we show that network coding can meet HRLLC requirements with the same or even fewer resources compared to existing 5G systems. Furthermore, we show that it reduces in-order delivery delay and ensures efficient resource utilization without degrading performance. The benefits of network coding increase when feedback is unreliable or delayed, making it ideal for applications in non-terrestrial networks (NTN), another major consideration in 6G standardization. Even though not directly highlighted in this paper, our results are applicable to multi-hop networks with D2D communication, and these benefits will continue to hold in more complex networks as well. The following considerations from standardization bodies would enable 6G systems to fully utilize the benefits of network coding in some of their use cases.

\subsubsection{Protocol considerations for network coding integration}

Our analysis clearly shows that network coding achieves much lower latency, particularly in-order delivery delay, compared to current 5G reliability mechanisms. This is a key requirement for XR/VR applications, and network coding is once again proving to be a transformative technology for such use cases. Previous efforts in the literature \cite{WiMAX_study,dias2023sliding} have demonstrated its benefits from higher layers, but in this paper, we showcase its advantages when combined with reliability mechanisms at lower layers. For our implementations of network coding,  we propose adding a virtual network coding layer at the intersection of PDCP and RLC layer, which can be activated with a coding flag in the packet header. As proposed in section \ref{sec:system_model}, the network coding layer can be considered as a sublayer of RLC layer and operates on the SDUs received at the RLC layer. While the network coding layer is activated, the HARQ transmissions in the MAC layer is reduced to one and the Unacknowledged Mode is used in the RLC layer to restrict ARQ retransmissions. With this approach, we ensure that there is no significant delay in bypassing these layers and follow the existing standards. We expect that 6G standardization efforts will continue to include similar options to bypass lower layer retransmissions. %This approach essentially replaces traditional reliability mechanisms with coding at a higher layer. However, the MAC layer frames still include LDPC coding as part of the first HARQ transmission.

3GPP standards would allow for backwards compatibility via per-bearer feature activation. A UE that supports network coding reliability handling could advertise this capability in its UE capability information. The gNB would recognize this capability and establish a network coding instance for each new UE joining the cell with that capability. UEs without this capability would operate under standard HARQ/ARQ mechanisms, thus allowing network coding to coexist with other forms of reliability in the same network. Proposed enhancements include capability signaling, coding parameter selection, and minor RLC/MAC header extensions to carry coefficient vectors and generation metadata. Such integration natively supports evolving XR, NTN, and IAB/WAB use cases targeted in Rel-18/19.

\subsubsection{Modularization of operations in standards}

While our approach essentially replaces traditional reliability mechanisms with coding at a higher layer, the MAC layer frames still include LDPC coding as part of the first HARQ transmission. It also occurs that PHY layer operations such as interleaving and MAC layer operations are tightly coupled in current 5G protocol stack implementations. This coupling restricts any attempt to replace the HARQ protocol at the MAC layer. 

As the industry becomes more flexible and open, making each layer more independent and allowing operational flexibility can significantly accelerate the adoption of newer technologies. In fact, standardization bodies can lead the way toward a more API-driven telecommunication architecture that allows proprietary coding schemes to operate seamlessly with existing infrastructure and hardware.

\subsubsection{Parallelization of processes and number of frames in flight}
Another important observation from our analysis regards the number of parallel processes or frames in flight. During the transition from 4G to 5G, the limitations on parallel HARQ processes and thus the number of frames in flight has become more stringent. The 5G systems limit the number of parallel HARQ processes to 16, considering that in terrestrial networks gNB to eNodeB communication can be completed within 16 transmissions. However, this introduced an unnecessary constraint on scenarios where feedback can wait or block acknowledgements can be enabled, such as network coding-based reliability mechanisms. The benefits of network coding scale with the number of frames in flight. In another aspect, if the round-trip time (RTT) in terms of packets or frames sent is higher, traditional systems suffer from increased in-order delivery delay and completion times. Network coding addresses this issue through its forward erasure correction capability, making it ideal for NTN applications, where larger delays between terrestrial and non-terrestrial devices are expected. Furthermore, the error rate in such applications is higher, and relying solely on feedback for error correction is not ideal.

However, 5G systems have an upper bound of 16 HARQ processes in parallel, which limits the number of frames in flight to 16. There have been discussions in 3GPP meetings about increasing this number, particularly in reference to NTNs. Our analysis shows that increasing the number of parallel processes would be beneficial not only for handling higher RTTs but also for maximizing the advantages of technologies like network coding. Furthermore, \cite{landon2024enhancing} presents additional techniques that can increase the number of frames in flight when using network coding.

\subsubsection{Use of network coding in specific use cases}
As mentioned earlier in this section, one of the major applications of network coding in 5G NR was discussed in the context of IAB, or more generally, multi-hop communications. The possibility of using recoding at intermediate nodes makes network coding a natural choice in such scenarios. A detailed study of network coding in IAB is presented in \cite{mao2024techniques,nikopour2023linear}. With discussions in standardization bodies now extending to Wireless Access Backhaul (WAB) - where an intermediate node can act as both receiver and sender with a full protocol stack - the use of network coding can provide significant benefits compared to traditional store-and-forward setups. This enables network coding based recoding at these intermediate nodes. This can be facilitated in higher layers as described in \cite{sundararajan2011network} or in lower layers as proposed in this paper.

Furthermore, coding can also be beneficial under duplex conditions, where both uplink and downlink signals at the intermediate node can be coded together to achieve better efficiency. Thus, applications such as WAB or multi-hop D2D communication benefit from network coding implementations. Applications that require a high traffic in both directionss, such as XR, would highly benefit from this capability. While sub-band full-duplex is considered for improved uplink throughput, network coding provides an alternative that can enhance per UE-gNB throughput for such use cases. It would be ideal for standardization bodies to consider such scenarios and advocate for intermediate nodes to have greater processing and memory capabilities.

In XR/VR use cases, the network coding approach can have even further impact. 3GPP TS 23.501 (Clause 5.37.5.2) \cite{3gpp.23.501} and TS 38.415 (Clause 6.5.3.9) \cite{3gpp.38.415} defines the new framework focusing on Protocol Data Unit (PDU) Set based framework for such applications where a group of PDUs are assigned a priority level and considered together for quality of service handling. The user plane protocol focused on PDU sets can benefit from network coding where high priority PDU sets can be coded together to ensure reliable transmission. Different priority levels can  be coded at different rates to ensure optimal bitrate and the same coding/decoding architecture can be used to process them. This approach aligns with the ongoing discussions in 3GPP SA2 focusing on XR applications.

\section{Conclusion}\label{sec:conclusion}
In this work, we presented a comparative study of network coding and existing 5G reliability mechanisms within a sliced network environment. We presented an analytical characterization of the goodput and delivery delays of HARQ and network coding erasure correction schemes. Through both theoretical characterizations and simulation-based evaluations, we demonstrated that network coding significantly improves in-order delivery delay and efficient resource utilization, as well as completion time and goodput in scenarios with higher RTT. Our findings suggest that integrating network coding into future wireless architectures can enable Hyper Reliable Low Latency Communication (HRLLC) with fewer resources and reduced latency, supporting emerging use cases such as XR, VR, and non-terrestrial networks.

Given its performance advantages, network coding should be strongly considered in ongoing 6G standardization efforts. In particular, we advocate for more modular layers in the protocol stack such that changes to one module do not affect another, to allow network coding to complement or replace HARQ/ARQ mechanisms, especially in slices demanding low latency and high reliability. Future work may include experimental validation on physical testbeds, exploration of sliding window coding schemes for further latency reduction, and an exploration of the potential effects of changes to come in 6G networks.

\bibliographystyle{IEEEtran}
\bibliography{references}

\end{document}